# SibRank: Signed Bipartite Network Analysis for Neighbor-based Collaborative Ranking

Bita Shams [a] and Saman Haratizadeh [a]
[a] University of Tehran, Faculty of New Science and Technology
Amir Abad, North Kargar Street, Tehran, Iran14395-1374

**Abstract**

Collaborative ranking is an emerging field of recommender systems that utilizes users' preference data rather than rating values. Unfortunately, neighbor-based collaborative ranking has gained little attention despite its more flexibility and justifiability. This paper proposes a novel framework, called SibRank that seeks to improve the state of the art neighbor-based collaborative ranking methods. SibRank represents users' preferences as a signed bipartite network, and finds similar users, through a novel personalized ranking algorithm in signed networks.

# 1 Introduction

Recommendation systems exploit users' historical data in order to suggest a small set of relevant services among a large volume of irrelevant ones. Traditionally, these systems seek to predict user's interests through gathering and prediction of ratings given by him to services or products. In recent years a variation of recommender systems, called collaborative ranking, has emerged that focuses on ranking data instead of ratings [1,2] to find the best items to recommend.

Rankings seem to be more informative and reliable source of information to reflect users' preferences and interests [2,3]: users' ratings commonly follow a baseline that is a function of time[4]. For instance, a user that assigns a rate of 4 to an average movie "A" now, might tend to assign 3 to that movie in the future. However, in both times, he probably would not prefer that average movie over an excellent movie "B". So, he persistently ranks "B" over "A" while the rates he gives to those movies may change over time. Furthermore, the goal of a recommender system is to find and recommend the most relevant items [3,5,6]. Therefore, it is more important to accurately predict the user's priorities rather than the absolute ratings he would give to items. For example, consider a case that a target user would rate the movies "A" and "B" with 4 and 5 respectively, that is r(A)=4 and r(B)=5. Suppose that some rate prediction algorithm, predicts ratings r(A)=2 and r(B)=3, while another algorithm predicts r(A)=*5 and r(B)=4*.  Although, in rate based framework, the latter one is considered more accurate in the sense of rate prediction, it is clear that the former one would correctly rank "B" higher than "A" and will probably make a better recommendation.

Recently, several researchers have approached recommendation from a ranking-oriented perspective. These approaches can be categorized into two groups: model-based collaborative ranking methods (MCR) and neighbor-based collaborative ranking methods (NCR). MCR techniques try to learn the latent factors of users and items while optimizing a ranking criterion (e.g. Normalized Discounted Cumulative Gain (NDCG))[1,5,7] while NCR algorithms exploit the concept of users' similarity to infer a ranking over the user's uncollected items[2,6,8–11]

Neighbor-based collaborative ranking has not been investigated as much as model-based ranking. The reason is possibly that it is not straightforward to calculate the similarity among users when each user's profile is a set of preferences [12] Most of the current NCR algorithm have exploited Kendall correlation that takes into account users' agreements and disagreements over pairwise comparisons [2,10,11]. However, Kendall measure, has originally been proposed to compare similarity between total orders (i.e. ranking over a set of distinct items)[12] Therefore, it is not directly applicable to recommender systems where each user has partially ranked a different subset of items. To resolve this issue, current NCR techniques calculate users' similarity based on their common comparisons and ignore the information available through the uncommon rankings. This approach suffers from some shortcomings: First, Kendall correlation does not take in to account the confidence or reliability of the source of information: the similarity calculated over a small set of common comparisons is not as reliable as one calculated over a large set of comparisons. Second, Kendall correlation between users with no common comparison is equal to zero, simply because this measure ignores all other available information in the data that can be used to discriminate between such users. These problem are more serious in in sparse datasets where users rarely have a common pairwise comparison. As illustrated in Fig.1a, in a sparse data set, Kendall correlation and its variants would be zero for more than 87% of pairs of users. Also, in average, more than 96% of the users among which the Kendall correlation is one, have only one common pairwise comparison (See Fig.1b). That clearly indicates that when one uses Kendall correlation for finding similar users, not only all users with no common comparison with the target user are ignored, but also the recommendation is highly affected by those users who have only one common pairwise comparison with the target user.

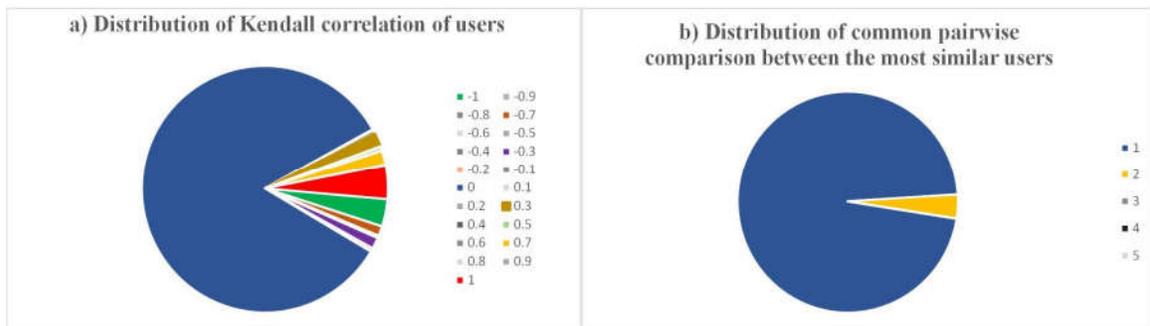

Figure 1. Statistical properties of Kendall Correlation in a sample dataset of movielense100K containing 10 ratings for each user. a) Distribution of Kendall correlation of users b) Distribution of the number of common pairwise preferences among users whose Kendall Correlation is one

This paper proposes a novel NCR framework, called SibRank, to resolve these issues. SibRank first construct a signed bipartite network, called SiBreNet, to represent users' preferences. Then, it exploits a novel similarity measure, called SRank, to calculate similarity among the target user and all other users in SiBreNet.

The main contributions of this paper can be summarized as below:

- To our knowledge, this is the first work that takes advantages of signed networks to capture the similarity among users' ranking. This paper represents ranking data as a signed bipartite graph to comprise different kinds of information about agreement and disagreement of users over their preferences.
- We propose a novel similarity measure in signed networks that reflects a global view of agreements and disagreements between users.
- We conducted set of experiments to assess the performance of SibRank. The results show some improvements compared to one of the best current algorithms called EigenRank.

## 2 Related works

## 2.1 Neighbor-based Collaborative Ranking

The first neighbor-based collaborative ranking algorithm was proposed in [2]. The paper introduced a general framework, called EigenRank, which is now widely followed by other NCR techniques[8–10,13]. It finds similar users to the target user based on users' preferences. Based on those neighboring users, EigenRank infers the preference matrix for the target user, using which it infers a ranking over the unseen items, and recommends the top ones to him. This framework is commonly used in all neighbor-based collaborative ranking algorithms with slight modifications; the main difference often lies in the similarity measures used by NCR techniques for finding similar users. EigenRank uses the Kendall Correlation measure and calculates the similarities considering the concordant and discordant pairwise preferences in the users' preferences. EduRank[13] modifies the Kendall measure by also considering compatible preference. The compatible case occurs when one user neither agrees nor disagrees another user over a distinct comparison. For example, consider three users *u*, *v*, *p* that all have compared two items (a, b) and the results are (*a>b*), (*a<b*), (*a=b*) for *u*, *v* and *p*, respectively. *u* and *v* disagrees over this comparison, while, *p's* verdict is compatible with those of both *a* and *b*. These versions of Kendall measure do not considers the importance of each preference for finding similar users to the target user. VSRank[8] introduces another measure, called weighted Kendall, to cover this issue. However, this approach, like the previous ones, still does not consider the reliability for the calculated similarities and also can't determine the proximity between users with no common pairwise comparisons. Additionally, it has been designed to exploit the rating data, and, consequently is not applicable in ranking-oriented systems with no ratings available.

In this paper, we propose a novel similarity measure that addresses the issues through personalized ranking in a signed network representing users' preferences. The proposed approach is applicable in all cases of preference data such as ratings, binary relevance feedbacks, and, pairwise preferences. Since, in our method, we model the data in the form of a graph, we will first present a brief review of the current available graph-based approaches for recommendation.

## 2.2 Graph-based recommendation

Graph structure provides effective tools to model and analyze direct and indirect relations among users. This has resulted in a large number of researches in graph-based recommendation algorithms. These algorithms commonly focus on binary data (i.e. like, dislike) and rarely pay attention to ratings. They use the graph structure to efficiently calculate some "closeness" value among users, items, or between users and items. Li and Chen formulated recommendation problem as calculating closeness of users and items in a bipartite graph. Anhui [14] investigated how to calculate similarity among users in a user-movie bipartite network. Another study [15] follows that approach in a tripartite graph containing users, items, and tags. The authors in [16] calculate closeness of all types of objects in a heterogeneous network. Xiang, et al. captured the long-term and short-term preferences of users in a tripartite graph of users, items, and sessions [17]. Yao, et al. have exploited a multi-layer graph to make context-aware recommendations [18]

To our knowledge, none of these algorithms focuses on modeling ranking and preference data through signed networks, and, consequently, SibRank is the first algorithm in the family of NCR algorithms that exploits signed networks to represent the ranking data and to calculate a similarity among users' priorities.

## 2.3. Personalized ranking in signed networks

Computing similarity of nodes to a particular node, termed as a personalized ranking of nodes, is one of the main challenging issues in social and information networks. Despite a large series of personalized ranking algorithms such as personalized PageRank or Katz similarity, few have focused on signed networks containing positive and negative relations. Ranking nodes in signed network have mainly been studied in the domain of trust/distrust relationships: Guha, et al.[19] extend the Katz similarity for signed networks through multiplication of signed adjacency network or considering one-step propagation of distrust. Additionally, a series of algorithms have been proposed through extending PageRank for signed networks. For example, EigenTrust [20] first removes the negative links from the network and then, ranks the nodes in the resulting network, so, a negative link between two nodes, is treated just like an absent link, while they represent totally different facts. PolarityRank [21] is another algorithm that classifies the nodes' labels through propagation of ranks from some predetermined representatives for both positive and negative groups. PageTrust [22] takes into account negative links through decreasing the number of random walkers at nodes with large negative links. Exponential ranking algorithm [23] is an extension of PageRank algorithms that modifies the transition probability matrix in the sense that the random walker follows positive links with higher probability than negative links. Similarly, some other ranking algorithms assume that nodes reaching from paths passing negative links should be ranked less than those reaching from paths passing positive links [24,25]. Essentially. these algorithms, first decompose the network to the negative and positive networks containing only negative and positive networks, respectively. Then, they define the ranking of each node based on the difference of its ranking in positive and negative networks. Another signed ranking algorithm, Troll-Trust [26], models the absolute value of nodes' reputation as a Bernoulli random variable that can be interpreted as how much one should trust a particular user.

Most of the mentioned algorithms extend PageRank considering *additive propagation* of distrust which refers to the social belief that" I don't respect someone not respected by someone that I don't respect" [19]. Although this idea is somehow reasonable in terms of distrust propagation, it is untenable for calculating users' similarities in signed bipartite preference networks. In this case, users who disagree the discordant preferences are probably as similar as those users who agree over concordant preferences. Therefore, we adopt a signed multiplicative propagation approach which follows another social belief, called social balance theory, that tells "enemies of my enemy and friends of my friend are friends, while, the enemies of my friend and friends of my enemy are enemies" [21].

## 3 SibRank Framework

SibRank is a novel NCR framework that seeks to improve the current state of the art algorithm through considering direct and indirect agreements and disagreements of users. It first maps the users' rankings to a **Si**gned **B**ipartite p**re**ference **Net**work, called SiBreNet). Then, it calculates the closeness of users to the target user, and, labels the most similar ones as neighbors. After that, it estimates the preference matrix of the target user based on his neighbors' verdicts, infers a total ranking for him and recommends the Top-N items.

## 3.1 Construction of SiBreNet

SiBreNet represents users' priorities using two types of nodes (i.e. users and preference) and two types of relations: positive and negative relations. A user is connected to a preference node with a positive edge if he agrees with that preference and is connected to a preference with a negative edge, if he disagrees with that preference. There is no edge between a user and a particular preference if his opinion is unknown about that preference. Figure 2 illustrates a schematic example of SiBreNet; *Jack* is connected with a positive edge to the preference node {*A>B*} because he ranks A over B. On the other hand, *Mike* believes that *B* is better than *A*, and, so *Mike* and {*A>B*} are connected with a negative link.

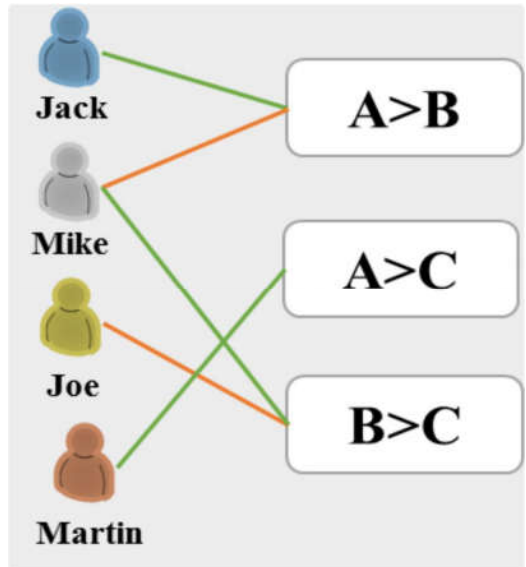

Figure 2: An example of SiBreNet in case of four users and three items. Red and green lines indicate negative and positive relations, receptively.

***Definition 1 (SiBreNet):*** *SiBreNet is a bipartite Graph (<V, E,W>) where V is the set of nodes and E is the set relations among nodes. V is composed of two subsets: user nodes U and preference nodes P; **V = U ∪ P**. Each edge (u,p) is labeled with a weight $\boldsymbol{w_{u,p}}$ that indicates the agreement ($\boldsymbol{w_{u,p} = 1}$) or disagreement ($\boldsymbol{w_{u,p} = -1}$) of user u to preference p.*

The pseudo-code of SiBreNet construction is shown in Algorithm.1. It initializes a bipartite graph including one node for each user (Line 1-3). Then, it makes an alphabetical order on items (Line 5) and creates the preference nodes (Line 6-8). For each pair of items $A, B$ the algorithm creates one preference node with regards to the order of items in the alphabetical order: if $A$ comes before $B$, it will add the preference node $\{A < B\}$, otherwise, node $\{B < A\}$ will be generated instead. Finally, for each pair of users and preferences $(u, p)$, it checks whether $u$ agrees with preference $p$, disagrees it, or has not judged the preference yet (Line 9-14). If $u$ agrees with $p$, then a positive link will be added between $u$ and $p$ (Line 11-12). If he disagrees $p$, a negative link will be generated (Line 13-14) and if $u$ does not judge $p$, or likes them equally, then no link will exist between them.

**Algorithm 1: Pseudo code for SiBreNet construction**

| | |
|---|---|
| **Input** | Set of users U, Set of items I, Preference database (P) |
| **Output** | Signed Bipartite Preference Network (G) |
| 1 | //Initializing user layer |
| 2 | For each user $u \in U$ |
| 3 |    Create a node $u$ in user layer |
| 4 | //Initializing preference layer |
| 5 | S= generate a permutation of items |
| 6 | for ($i=1;i<length(s);i++$) |
| 7 |    for($j=(i+1);j<=length(s);j++$) |
| 8 |      create a preference node indicating ($S(i)<S(j)$) in preference layer |
| 9 | for each user $u \in U$ |
| 10 |    for each $p \in P$ |
| 11 |      if $u$ agrees $p$ |
| 12 |        create a positive edge between $u$ and $p$ |
| 13 |      else if $u$ disagrees $p$ |
| 14 |        create a negative edge between $u$ and $p$ |

## 3.2 SRank: personalized ranking in signed networks

We present an algorithm called SRank that adopts signed multiplicative rank propagation [21] for similarity calculation in recommender systems. SRank exploits a personalized random walk in an undirected signed graph for calculating similarities among nodes, and is based on some ideas from social balance theory. Social balance theory refers to the fact that for a person, friends of friends, and, enemies of enemies are friends, while, friends of enemies, and, enemies of friends are enemies [27,28].

**Definition 2: (Positive and negative random walk):** *A random walk* $P = <v_1, v_2, \ldots, v_m>$, *passing nodes* $v_1, v_2, \ldots, v_m$ *is called a **positive random walk** with length m if* $\prod_{i=1}^{i=m-1} w_{v_i,v_{i+1}} = 1$. *Also, a random walk* $Q = <u_1, u_2, \ldots, u_n>$ , *passing nodes* $u_1, u_2, \ldots, u_n$ *is called a **negative random walk** with length n if* $\prod_{i=1}^{i=n-1} w_{u_i,u_{i+1}} = 1$

The basic idea behind SRank is that a positive random walk from a source node "s" to a destination node "d" is a sign of similarity between "s" and "d" while existence of a negative walk from "s" reaching "d" is a clue for the dissimilarity between them.

Formally, for a given signed graph $G=<V,E, W>$, a random walker will surf the graph according to the transition matrix *T*. An element $T_{i,j}$ in matrix T indicates the probability that a random walker at node *i* will move to node *j* and is defined by Eq.1

$$T_{i,j} = \frac{|w_{i,j}|}{\sum_{k=1}^{|V|} |w_{k,j}|} \tag{1}$$

where $w_{i,j}$ is the weight of edge between nodes *i* and *j*. In fact, the random walker at node *i* will move to one of its neighbors uniformly with probability of $1/d_i$ where $d_i$ is the degree of node "*i*".

Since SRank requires to distinguish positive and negative random walks, it decomposes the transition matrix to the positive and negative components that are obtained through Eq.2 and Eq.3, respectively.

$$T_{i,j}^+ = \begin{cases} T_{i,j}, & w_{i,j} = 1 \\ 0, & O.W \end{cases} \tag{2}$$

$$T_{i,j}^- = \begin{cases} T_{i,j}, & w_{i,j} = \text{-}1 \\ 0, & O.W \end{cases} \tag{3}$$

Positive and negative transition matrices indicate the probability that a random walker at node *i* will move to one of its positive and negative neighbors, respectively. According to these matrices, we can track positive and negative random walks reaching each node. Given these transition matrices, the positive SRank ($S^+$) and negative SRank ($S^-$), can be obtained through Eq.4 and Eq.5, respectively:

$$S^+ = \alpha(T^+S^+ + T^-S^-) + (1-\alpha)q \tag{4}$$
$$S^- = \alpha(T^-S^+ + T^+S^-) \tag{5}$$

where α is a damping factor usually set to 0.85, and *q* is the personalization vector obtained by Eq.6.

$$q_i = \begin{cases} 1, & i = u \\ 0, & o.w \end{cases} \qquad (6)$$

where $u$ is the query node.

Given the positive and negative SRank values of nodes, their overall SRank of node $v$ is calculated according to Eq.7.

$$S(v) = \frac{S^+(v) - S^-(v)}{S^+(v) + S^-(v)} \qquad (7)$$

Where $S^+(v)$ and $S^-(v)$ are positive and negative SRank of node $v$ respectively.

Algorithm 2 depicts the pseudo-code of SRank algorithm. It first conducts an initialization process by constructing the adjacency matrix, the transition matrices, and the personalization vector (Line 1-3). Then, it initializes positive and negative SRank values (Line 4). Note that positive and negative SRank values represent probabilities of reaching nodes with random walks through positive or negative paths. Therefore, they should be normalized to their summation (Line 5-8). After this phase, the algorithm iterates the process of updating positive and negative SRank values until convergence (Line 10-13). Finally, it computes the overall rank of nodes from the target user's perspective (Line 14)

**Algorithm 2: Pseudo code of SRank**

| | |
|---|---|
| **Input** | Signed Bipartite preference Network $G <V, E>$, target user $u$, and damping factor $\alpha$ |
| **Output** | Personalized ranking of nodes from the perspective of the target user |
| 1 | Initialize a $M$ through Eq.1 |
| 2 | Compute $T$, $T^+$, and $T^-$ according to Eq.1, Eq.2, and Eq.3 |
| 3 | Initialize personalization vector $q$ according to Eq.6 |
| 4 | Random initialization of SRank vectors $(s_0^-)$ and $(s_0^+)$) |
| 5 | $sn = sum(s_0^-)$ |
| 6 | $sp = sum(s_0^+)$ |
| 7 | $s_0^- = \dfrac{s_0^-}{sp + sn}$ |
| 8 | $s_0^+ = \dfrac{s_0^+}{sp + sn}$ |
| 9 | t=0; |
| 10 | Repeat until Converge: |
| 11 | $\quad S_{t+1}^+ = \alpha(S_t^+ T^+ + S_t^- T^-) + (1 - \alpha)q$ |
| 12 | $\quad S_{t+1}^- = \alpha(S_t^+ T^- + S_t^- T^+)$ |
| 13 | $\quad$ t=t+1; |
| 14 | compute SRank of nodes according to Eq.7 |
| 15 | return SRank of nodes |

### 3.2.2 Properties of SRank

**Property.1:** *Positive and negative SRank values will converge.*

**Proof**. Using similar ideas from application of signed multiplicative rank propagation in nodes' classification[21], Eq.5 and Eq.6 can be rewritten as Eq.8

$$\begin{bmatrix} S^+ \\ S^- \end{bmatrix} = \alpha \begin{bmatrix} T^+ & T^- \\ T^- & T^+ \end{bmatrix} \begin{bmatrix} S^+ \\ S^- \end{bmatrix} + (1-\alpha) \begin{bmatrix} q \\ z \end{bmatrix} \tag{8}$$

where $q$ is the personalization vector, z is a ($|V|\times 1$) vector of all zeroes, and $\alpha$ is the damping factor. Therefore, positive and negative SRank can be calculated through computing personalized PageRank of the graph with transition matrix of $D$ and personalization vector that are defined through Eq.9 and Eq.10, respectively:

$$D = \begin{bmatrix} T^+ & T^- \\ T^- & T^+ \end{bmatrix} \tag{9}$$

$$r = \begin{bmatrix} q \\ z \end{bmatrix} \tag{10}$$

Personalized PageRank of the transformed graph will be in the form of $\begin{bmatrix} S^+ \\ S^- \end{bmatrix}$ and can be easily decomposed into positive and negative SRank values. Note that $D$ is a column stochastic matrix, since the sum of elements in each of its column is 1. Also we know that $\|p\|_1=1$ and $\|r\|_1=1$. Therefore, personalized PageRank algorithm will converge to the eigenvector corresponding to the largest right eigenvalue of matrix $(\alpha D + (1-\alpha)r\mathbf{1})$ where **1** is a vector of all ones[29].

**Property.2:** *Positive and negative SRank of a signed network sums to personalized PageRank of its unsigned form.*

**Proof.** Considering summation of positive and negative SRank values:

$$S^+ + S^- = \alpha(T^+S^+ + T^-S^-) + (1-\alpha)q + \alpha(T^-S^+ + T^+S^-)$$

Therefore,

$$S^+ + S^- = \alpha(T^+S^+ + T^-S^- + T^-S^+ + T^+S^-) + (1-\alpha)q$$

Since $T = T^+ + T^-$ we have

$$S^+ + S^- = \alpha T(S^+ + S^-) + (1-\alpha)q$$

So, $(S^+ + S^-)$ is equal to personalized PageRank of graph with transition matrix $T$, and, positive and negative SRank can be considered as decomposition of the personalized PageRank to elements representing the positive and negative random walks.

### 3.2.3 Algorithm Explanation

SRank estimates the agreements and disagreements between the target user and other users using the comprehensive information provided by the graph structure. The importance of this global view for finding similar users in a graph can be illustrated through an example: In Figure 2, *Jack* is connected with a negative path of length 2 to *Mike* and with a positive path of length 4 to *Joe* while he has no path to *Martin*. This implies that *Jack* is more likely to be similar with *Joe* than *Mike* and *Martin*. That makes sense because *Jack* and *Joe* have a common neighbor, *Mike,* with whom they both disagree, while there exists no clue about similarity or dissimilarity between *Jack* and *Martin*. These facts are simply neglected by local approaches such as Kendall measure: using Kendall measure similarity of *Jack* to *Mark*, *Joe* and *Martin* will be -1, 0 and 0, respectively. Therefore, it seems more informative to see the agreements and disagreements among users' preferences from a global perspective. SRank models the direct and indirect agreements and disagreements using the concept of negative and positive random walks introduced in Definition 2.

To make these concepts more clear, see the example shown in Figure 3. Suppose that we seek to calculate the overall agreement and disagreement between *Kevin* and *Matt*. They are connected to each other by positive or negative paths, passing through a subset of *known preferences of Matt.* *These preferences* can be decomposed into two sets: concordant preferences *(i.e. A>c and B>C)* and discordant ones *(i.e.A>B)*. A Positive random walk from *Kevin* to *Matt* could be recursively defined as either

- A positive random walk from *Kevin* to *Matt* passing from a concordant preferences of *Matt*
  Or
- A negative random walk from *Kevin* to *Matt* passing from a discordant preferences of *Matt*

Also, A Negative random walk from *Kevin* to *Matt* could be recursively defined as either:

- A negative random walk from *Kevin* to *Matt* passing from a concordant preferences of *Matt*
  Or
- A positive random walk from *Kevin* to *Matt* passing from a discordant preferences of *Matt*

In this example, *Kevin* is connected to *Matt* through 3 positive paths and 5 negative ones.

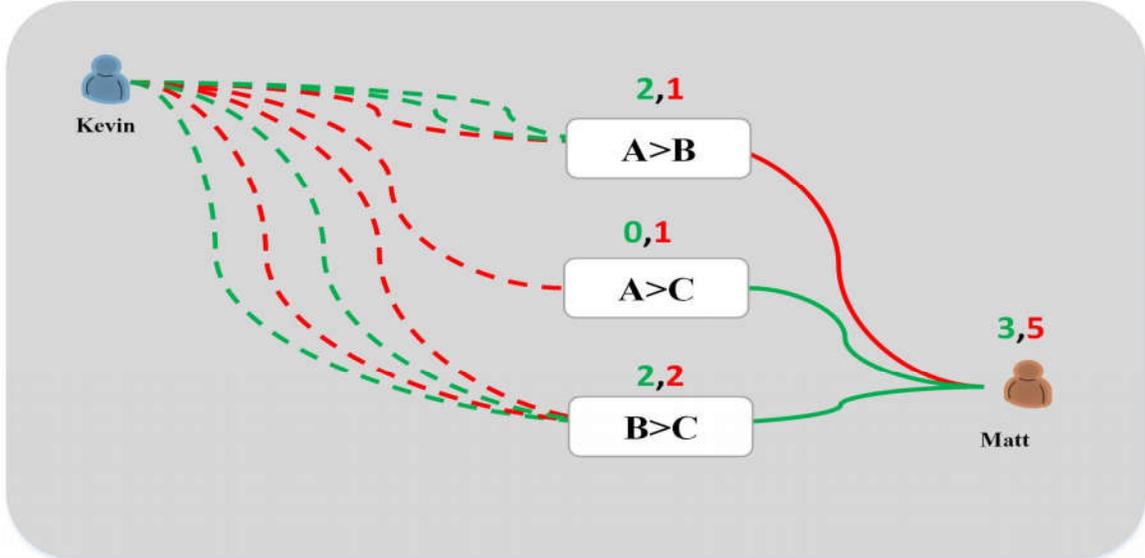

Figure 3: Green edges indicates positive relations, while, red edges indicates negative relations. Solid and dash lines differentiate direct and indirect relationships among users.

SRank extends this idea using a personalized random walk process that considers the length of walks; It keeps a positive and a negative SRank value for each node, that reflects the total agreement and disagreement of that node with the source node of the walk. A random walker starting from a source node, in each node "n", follows a random link with probability $\alpha$ and restart the walk from source node with probability $1-\alpha$. If he follows a positive link, the positive and negative SRank values of destination node are updated by positive and negative SRank values of the node "n", respectively. On the contrary, if he follows a negative link, the positive and negative SRank values of the destination are respectively updated using the negative and positive SRank values of node "n". In the case of a random restart, the random walker jumps back to the target node and the positive SRank of the target node is increased by $1-\alpha$. The stationary distribution of positive and negative SRank values are calculated through iteratively computing Eq.5 and Eq.6 (As depicted in Algorithm 2, Line 10-14).

### 3.2.4 Computational complexity

According to Property.1, SRank can be computed at the computational complexity of computing personalized PageRank with $2|V|$ vertexes, and, $2|E|$ edges. Therefore, its computational complexity is equal to $2E$ for sparse networks. The number of edges is equal to the $W$ where $W$ is the number of preferences assigned by all users. As a result, SRank computes the similarity between the target users and all others users in $O(W)$. Clearly $W$ is at most $MN^2$, so SRank's time complexity is in $O(MN^2)$ although $W$ is usually much smaller than $MN^2$.

It is worth noting that Kendall correlation of two users is computed at computational complexity of $O(N^2)$ in a system containing M users and N items. So, calculation time needed for finding the similarities between the target users and other users is in $O(MN^2)$ for the Kendall measure.

## 3.3 Ranking inference

To infer a total ranking, SibRank first estimates the preference matrix of the target user based on his neighbors through Eq.11

$$\hat{\psi}_{i,j}^u = \frac{\sum_{v \in neighborhood(u)} S_u(v)\psi_{i,j}^v}{\sum_{v \in neighborhood(u)} |S_u(v)|} \quad (11)$$

where $\psi_{i,j}^v$ is the real preference of $v$ on $(i,j)$ that is defined as:

$$\psi_{i,j}^v = f(x) = \begin{cases} 1, & \text{if } v \text{ prefers } j \text{ over } i \\ 1, & \text{if } v \text{ prefers } i \text{ over } j \\ 0, & \text{if } v's \text{ preference about } i \text{ and } j \text{ is uknown} \end{cases} \quad (12)$$

The next step is to infer a total ranking by aggregating the elements of the estimated preference matrix. To do this, SibRank exploits exponential ranking of items in the signed graph with adjacency matrix of $\hat{\psi}$. Exponential ranking calculates the ranking of item through PageRank calculation of a weighted graph with adjacency matrix of $\varphi$ (i.e. transition probability matrix) obtained by Eq.13

$$\varphi(i,j) = \frac{e^{\hat{\psi}_{i,j}^u}}{\sum_{k \in I} e^{\hat{\psi}_{i,k}^u}} \quad (13)$$

Finally, SibRank recommends the $TopN$ items that have the highest ranks. It is worth noting that the ranking inference procedure follows the social respect theory and additive rank propagation; for instance, if $x$ is less favorable than y (i.e. $\hat{\psi}_{y,x}^u = 1$) and $y$ is less favorable than $z$ (i.e. $\hat{\psi}_{z,y}^u = 1$), then one can expect that x is much less favorable than z; hence, $x$ should be ranked lower than $z$. Therefore, signed multiplicative rank propagation is not a wise choice for ranking inference since it will mistakenly propagate a positive rank from $z$ to $x$.

### 3.3.1 Computational complexity of the ranking inference

The ranking inference is composed of two steps: estimating preference matrix, and preference aggregation. Estimating preference matrix is computed at computational complexity of $O(KN^2)$ where k is the number of neighbors, and, $N$ is the number of items. Additionally, computational complexity of preference aggregation is equal to number of non-zero elements in preference matrix, and, consequently is less than $(N^2)$. To summarize, computational complexity of ranking inference is $O(KN^2)$. It should be noted that to our knowledge, computational complexity of this phase is not less than $O(KN^2)$ for any of available algorithms (e.g. EigenRank[2], Cares[10], VSRank [8], etc.)

## 3.4. Algorithm summarization

Algorithm.3 summarizes the overall framework of SibRank that consists of five steps: SiBreNet construction of preference data, calculating SRank of the target user *u*, finding the neighbors, ranking inference, and Top-k recommendation.

**Algorithm 3: Pseudo code of SibRank Framework**

| | |
|---|---|
| **Input** | Set of users U, Set of items I, Preference database (P), target user *u*, size of neighborhood *k*, length of recommendation list *N* |
| **Output** | Top-N recommendation list (*recomList*) |
| 1 | SiBreNet=SiBreNet-construction (*P, U, I*) |
| 2 | $S = $ SRank(SiBreNet, u, $\alpha$) |
| 3 | Neighbors= the set of *k* users with highest similarity |
| 4 | r= Infer ranking of users based on preference of Neighbors |
| 5 | recomList= Top-N items in *r* |
| 6 | return recomList |

# 4 Experiments and Result

We conducted all experiments on publicly available movie dataset, Movielens-100K, that is widely used in related works [3,6,8,11]: The dataset contains 943 users, 1,682 movies, and 100,000 ratings. Since general NCR techniques are designed for pairwise preference, we convert the rating information into a set of pairwise comparisons. To do so, we create a preference data <*u,i,j*> if the user *u* rated item *i* higher than item *j*.

To evaluate the performance of SibRank, we follow a standard protocol in collaborative ranking literature [1,3,6,11,30]. We assess the performance of the algorithms under different sizes of user profiles, i.e. the number of user's verdicts. For each user, a fixed number (*T*) of ratings will be selected for training and his remaining ratings will be left for test. Generally, collaborative ranking techniques are evaluated with 10, 30, and, 50 ratings in the training set for each user (i.e. T=10, 30, 50). For each *T*, we make sure that each user has at least 10 items in his test set. The reason is that the algorithms are assessed based on their Top3, Top5, and, Top10 recommendations. For each value of *T*, we generate 5 variants of training sets via random sampling, and, report the average of the performance on their corresponding test sets.

## 4.1 Baseline algorithm

We compared SibRank to EigenRank[2], the state of the art algorithm in the area of neighbor-based collaborative ranking. EigenRank calculates the users' similarity through Kendall correlation; estimates a preference matrix through linear combination of the neighbors preferences; and infers a total ranking from it using a random walk method on a graph constructed based on the preference matrix. Its computational complexity is $O(MN^2 + KN^2 + N^2)$ for *M* users, *N* items, and neighborhood size of *K*.

## 4.2 Evaluation Metrics

Following the evaluation protocol for collaborative ranking techniques, we compared the performance of SibRank and EigenRank based on their top-N recommendations: A recommendation algorithm is a good one if it suggests items that have the highest ratings in the test set. Normalized Cumulative Discounted Gain (NDCG), precision and recall are evaluation metrics that are widely used for assessing top-k recommendations. NDCG for the top-N recommendation for a user $u$ can be obtained by:

$$N@TopN = \frac{1}{\beta_u} \sum_{i=1}^{TopN} \frac{2^{r_i^u}-1}{\log_2^{i+1}} \tag{14}$$

Where $TopN$ is the length of recommendation list, $r_i^u$ is the rating given by user "$u$" to the $i\ t$ item in the recommendation list, and $\beta_u$ is the normalization factor to ensure that NDCG of the ideal recommendations to user "$u$" is equal to 1.

Precision and recall assess the $TopN$ recommendation based on the binary relevance of the recommended items. Relevant items are those items given higher rates than the average rating of the target user. $P@TopN$ and $R@TopN$ show the precision and recall of the $TopN$ recommendation:

$$P@TopN = \frac{N_{rp}}{N_r} \tag{15}$$

$$R@TopN = \frac{N_{rp}}{N_p} \tag{16}$$

where $N_p$ is the total number of relevant items, $N_r$ is the number of recommended items, and, $N_{rp}$ is the number of relevant items in the recommendation list

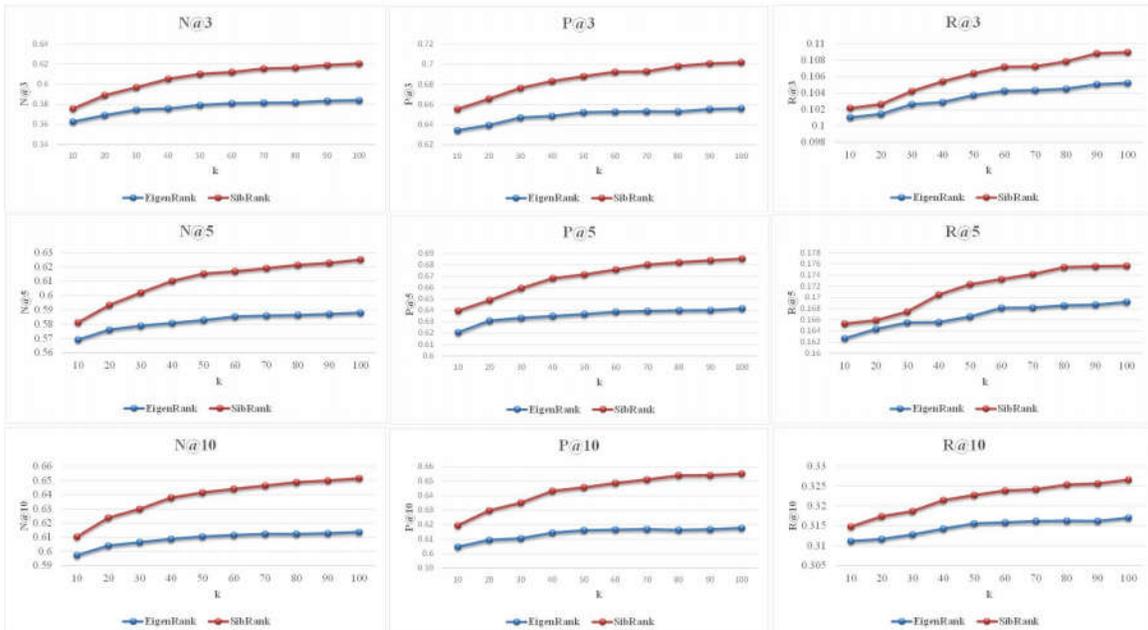

Figure 4: Performance comparison of EigenRank and SibRank when the length of user profile is 10

## 4.3 Results

We compared the performance of EigenRank and SibRank with different numbers of neighbors, from 10 neighbors to 100 neighbors. The results are depicted in Fig.4, Fig.5, and Fig.6. We analyzed the performance of SibRank and EigenRank from two perspectives: the length of users' profiles that is the number of users' preference in training data, and, the neighborhood size. As can be seen, the length of users' profiles has a great impact on the performance of both algorithms. For instance, for neighborhoods of size 100, increasing the length of the profiles from 10 (Fig.4) to 50 (Fig.6) causes a raise in NDCG@5 from 58% to 69% for EigenRank, and from 62% to 71% for SibRank. Clearly, strong profiles of users helps the algorithms to more accurately estimate the similarities between users. Results show that SibRank is more effective than EigenRank for all profile sizes.

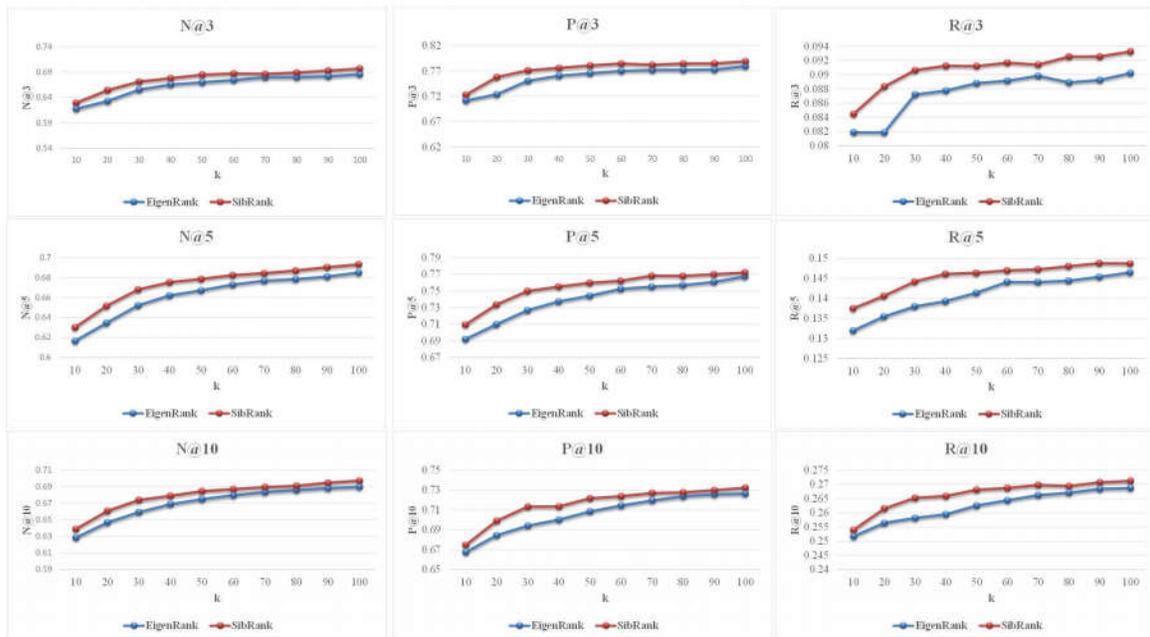

Figure 5: Performance comparison of EigenRank and SibRank when the length of user profile is 30

Our experimental results can be summarized as follows:
- SibRank outperforms EigenRank in all evaluation conditions. To be more precise, SibRank exhibits the improvement of 2%- 5% in different sizes of neighborhoods and users' profiles.

- SibRank significantly outperforms EigenRank using larger neighborhoods in sparse data sets (i.e. profile size of 10). For instance, in case of neighborhood size of 100, SibRank achieves N@10 of 65%, P@10 of 65%, and R@10 of 32%, while, performance of EigenRank does not exceed 62% for N@10, 61% for P@10, and, 31% for R@10 (See Fig.4).

- SibRank significantly outperforms EigenRank using small neighborhoods when large user profiles (i.e. 50) are available. SibRank shows improvement of 3% in term of N@3, 6% in term of P@3, and 0.6% in term of R@3 in a dataset with profile size of 50 and neighborhood size of 10 (See Fig.6).

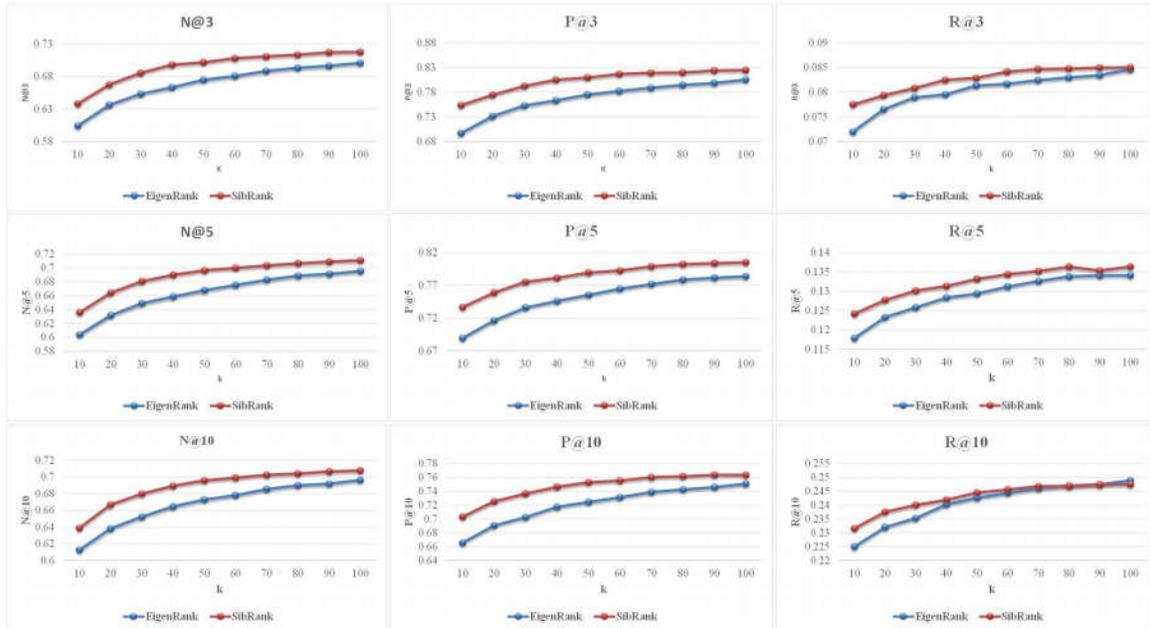

Figure 6: Performance comparison of EigenRank and SibRank when the length of user profile is 50

## 5 Discussion

As mentioned before, current neighbor-based collaborative ranking methods suffer from two major shortcomings: First, the similarity between all users with no common pairwise preference is equal to zero. Second, they do not consider the confidence of similarities. Here, we investigate how SRank and consequently, SibRank address these shortcomings and improve performance of neighbor-based collaborative ranking.

Figure 7 illustrates the distribution of users' similarities for EigenRank and SibRank for different sizes of user profiles. For profile size of 10, EigenRank cannot find any similarity among more than 85% pairs of users who have no common comparison. On the other hand, SibRank can efficiently propagate users' similarities through SiBreNet, and, accordingly, less than 5% of users have similarity around zero. Although increasing the size of users' profiles declines the number of users with no common comparison, EigenRank still cannot calculate similarity for more than 20% pairs of users in a data set containing users with profile size of 50.

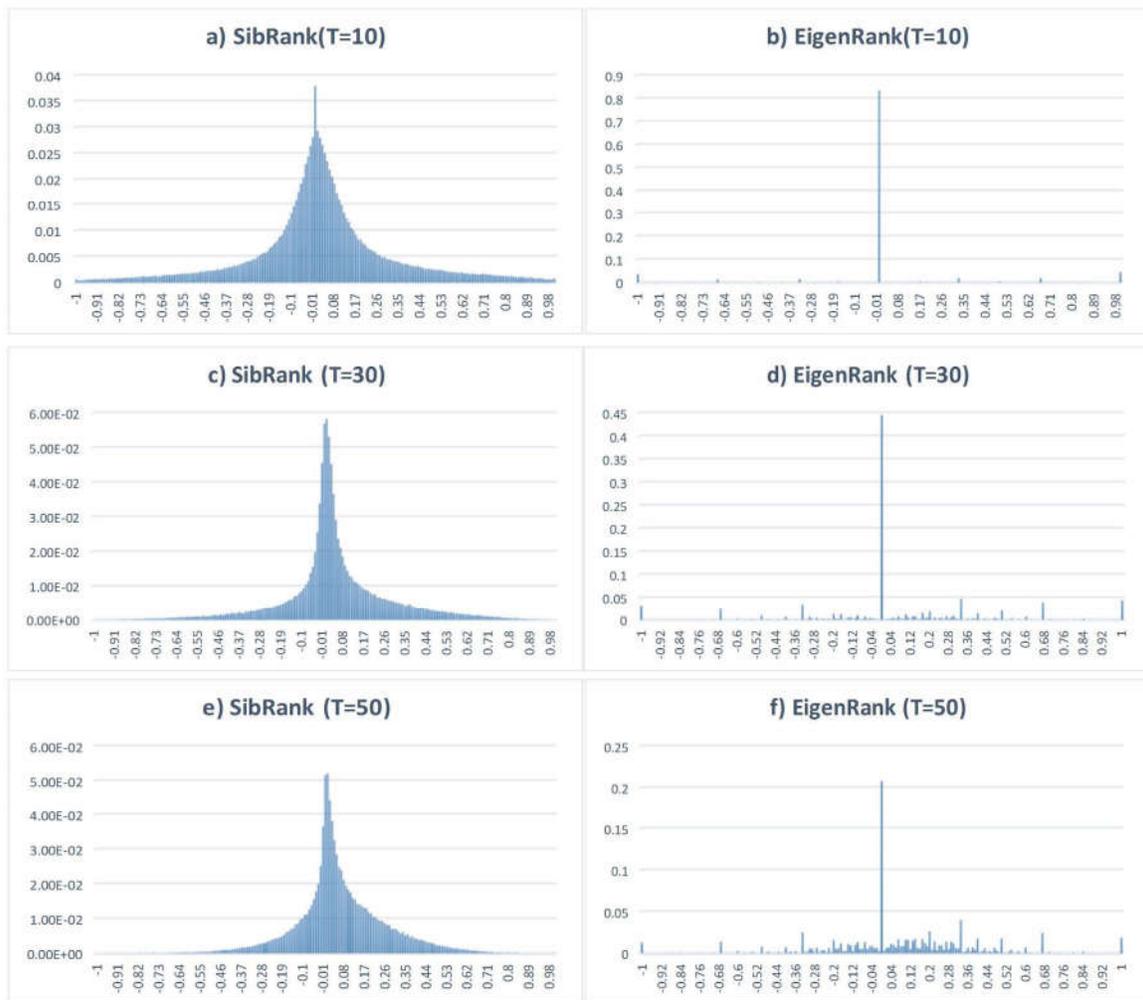

Figure 7. Comparison of EigenRank and SibRank in terms of distribution of user similarities among users for different sizes of users' profiles (T)

Another problem associated with current similarity measures is that they neglect the reliability or confidence of similarity measures. Figure 8 shows the similarity between three random users and their 100 nearest neighbors. As illustrated in Fig.8, EigenRank finds a large number of users with equal similarity to the target user, without considering the confidence of the calculated similarity. In such a case, EigenRank has to randomly select a number of users from this set of equally similar ones. In addition to this flaw, EigenRank also prioritizes neighbors solely based on their similarities, without considering how sure it is about the calculated similarities. To make the concept clear, see Fig.1b and Figure.8. As depicted in Fig.8, EigenRank Typically finds more than 20 users with Kendall correlation of one to the target user. On the other hand, Fig.1b shows that more than 96% of users with Kendall Correlation of one, have only one common pairwise preference. It evidently shows that EigenRank makes recommendation to the target user based on the preferences of the users who agreed over one pairwise comparison, while intuitively, a user with a larger number of agreements would be a better candidate for being a neighbor. In situations in which large user profiles are available, users have a higher number of pairwise comparisons, and so, for a target user, there are potentially many neighbors with more than one common comparisons. For example, in the data set containing user profiles of size of 50, each user has a small number (i.e. about 20) of users with agreement over one common pairwise comparison, and, a large number of users with higher number of agreements, but lower agreement ratio (e.g. Kendall correlation of 0.8 over 100 common pairwise comparison). However, EigenRank would prioritize the former group over the latter one, and if we are looking for small neighborhoods, only members of the former group will be considered by EigenRank as neighbors. Therefore, it could not efficiently recommend until the neighborhood size is big enough to have room for the members of second group as shown in Fig.6. That explains why SibRank significantly outperforms EigenRank when using small neighborhoods in denser data sets. It should be emphasized that good recommendation based on a smaller neighborhood sizes $K$, will reduce the running time in large data sets since the time complexity of both algorithms is O($MN^2 + KN^2 + N^2$).

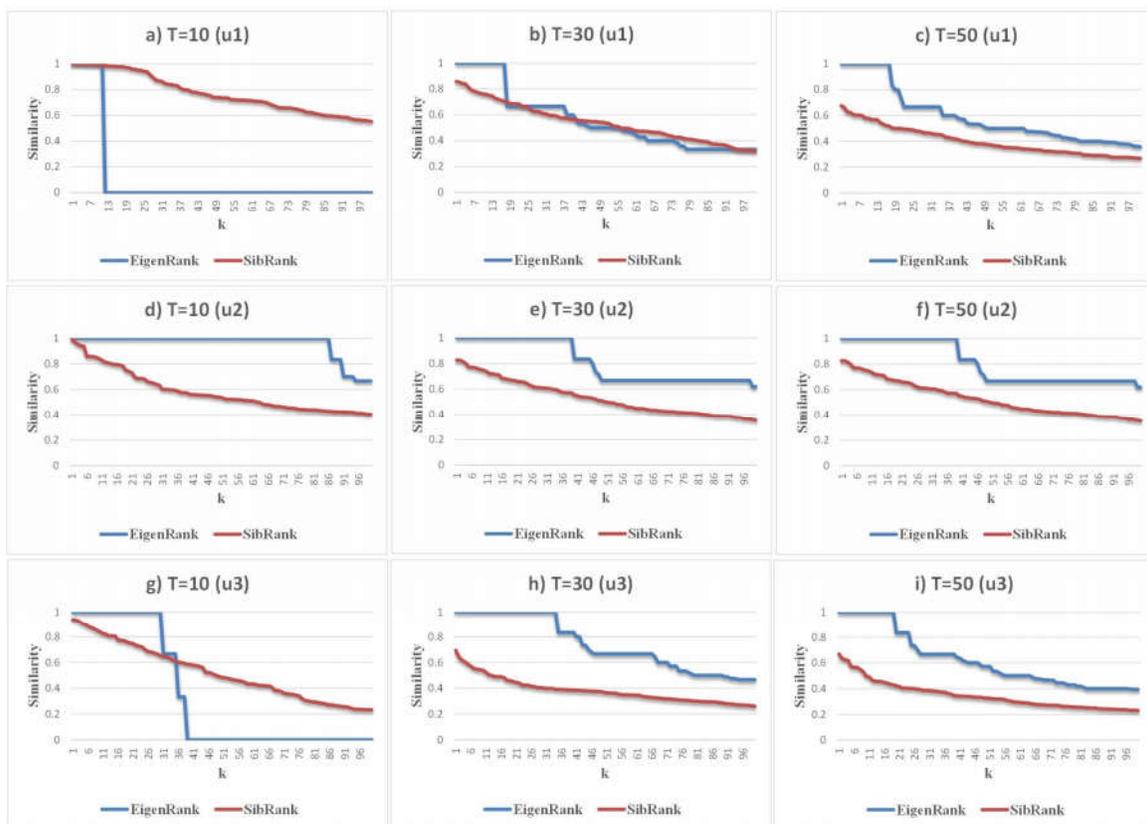

Figure. 8. Comparison EigenRank and SibRank in terms of Similarity between three random target user and their kth-nearest neighbors in case of different size of users' profiles (T).

If we have small profiles for users, the common pairwise comparison among users are rare. Therefore, there are a large number of users with zero Kendall correlation to the target user. So, increasing the neighborhood size necessitates EigenRank to expand the neighborhood to a random subset of those users with zero Kendall correlation (See Fig.8a and Fig.8g). Consequently, for users with small profiles, it cannot efficiently increase its performance by considering larger neighborhoods, as can be seen in Fig.4.

Generally, SibRank is more efficient in all conditions since it takes advantage of a graph structure to take into account the users' agreements and disagreements in addition to number and importance of pairwise comparisons in the similarity calculation of users.

# 6 Conclusion

In this paper, we proposed a novel framework, called SibRank, for collaborative ranking to improve recommendations' accuracy. SibRank introduces a network representation of ranking-oriented recommender systems, called SiBreNet, and exploits that for calculating users' similarities. To calculate users' similarities, we presented SRank, a personalized ranking algorithm for signed networks. The main difference between SRank and local similarity measures like Kendall, is that SRank calculates indirect similarities too, by modeling the data in a graph structure and calculating similarities among its nodes, by analyzing the positive and negative paths with different lengths. It is also different from other personalized ranking measures [22–24,31] as it models the social balance theory, to take into account both agreements and disagreements among users when calculating the similarities.

After finding similarities between a target user and other users, SibRank exploits a random-walk aggregation model to infer a total ranking for the target user based on his neighbors' preferences modeled by a preference matrix. Experimental results showed that SibRank achieves a better performance compared to EigenRank.

There are several interesting directions for extending the current work, such as redesigning the network structure to make a better representation of users' preferences, using other network-oriented ranking and similarity measures for recommendations based on SiBreNet, and adjusting SibRank for the domain of rating-oriented recommendations.